\documentclass[preprint,showpacs,preprintnumbers,amsmath,amssymb]{revtex4}
\usepackage{graphicx}

\begin{document}
\title{Modeling the Q-cycle mechanism  of transmembrane energy conversion}

\author{Anatoly Yu. Smirnov$^{1,2}$ and Franco Nori$^{1,2}$}
\address{$^1$ Advanced Science  Institute, RIKEN, Wako-shi, Saitama, 351-0198, Japan}
\address{$^2$ Physics Department, The University of Michigan, Ann Arbor, MI 48109-1040, USA}

\begin{abstract}
The Q-cycle mechanism plays an important role in the conversion of the redox energy into the energy of the proton electrochemical gradient
across the biomembrane. The bifurcated electron transfer reaction, which is built into this mechanism, recycles one electron, thus, allowing to
translocate two protons per one electron moving to the high-potential redox chain. We study a kinetic model of the Q-cycle mechanism in an
artificial system which mimics the $bf$ complex of plants and cyanobacteria in the regime of ferredoxin-dependent cyclic electron flow. Using
methods of condensed matter physics, we derive a set of master equations and describe a time sequence of electron and proton transfer reactions
in the complex. We find energetic conditions when the bifurcation of the electron pathways at the positive side of the membrane occurs
naturally, without any additional gates. For reasonable parameter values, we show that this system is able to translocate more than 1.8 protons,
on average, per one electron, with a thermodynamic efficiency of the order of 32\% or higher.
\end{abstract}

\maketitle

\section{ Introduction}
The energy produced by a biological system or by an artificial device must often be converted into a more stable form \cite{LaVan06,Kamat07}.
The thermodynamic efficiency and the quantum yield of this process is of prime importance for the overall performance of the energy transducer.
This transducer consumes the energy of input particles, which move energetically downhill or just disappear in the process, and transfer this
energy to another kind of particles moving energetically uphill. Here, the quantum yield ($QY$) is defined as the number of particles at the
output of the energy transducer divided by the number of input particles. The efficiency of such device can be higher when the quantum yield is
more than one, i.e., when a single input particle creates many output carriers.

The generation of two or more electron-hole pairs (excitons) by a single high-energy photon \cite{Nozik08} was observed in semiconductor
nanocrystals \cite{Schaller04}. A similar situation takes place in the $bc_1$ complex embedded into the inner mitochondrial membrane  as well as
in the related complex $bf$, which mediates the electron transfer between the Photosystem II (PS II) and the Photosystem I (PS I) in the
thylakoid membranes of plants and cyanobacteria \cite{Nicholls02}. According to the generally accepted Q-cycle mechanism
\cite{Mitch76,Osyczka05,Crofts06}, the transfer of \emph{two} electrons from a plastoquinol molecule PQH$_2$ to plastocyanin (in $bf$ complexes)
is accompanied by an energetically-uphill translocation of \emph{four} protons from the negative (N) to the positive (P) side of the membrane,
resulting in a quantum yield $QY = 2$.

We note that within the standard redox loop mechanism (see \cite{Richardson02} and references therein), only two protons are transferred in
parallel with the transfer of two electrons, implying that the quantum yield  is equal to one. Hereafter, we primarily concentrate on the $bf$
complex as a biological counterpart of our artificial system. Despite numerous studies \cite{Osyczka04,Cape06,Armen10,Cramer11}, the physical
mechanism of the Q-cycle in $bc_1$ and $bf$ complexes is not completely understood.

In this work we analyze a simple model (see Fig.~1) mimicking the main features of the Q-cycle in the $bf$ complex in the regime of
ferredoxin-dependent cyclic electron flow \cite{Joliot06,Shikanai07}. In this regime, electrons cycle between the PS I and the $bf$ complex,
which are electronically connected by a pool of ferredoxin molecules (on the N-side of the membrane) and by a pool of plastocyanin molecules on
the lumenal (P) side of the membrane. We treat these two pools as a source (S) and drain (D) electron reservoirs coupled to the electron-binding
sites A and B, respectively. Besides the sites A and B, the membrane-embedded central complex is comprised of sites L and H, which correspond to
hemes $b_L$ and $b_H$ of the complex $bf$. The sites A and H are assumed to be electronically decoupled as well as the sites B and L. A mobile
shuttle $Q$ (an analog of a plastoquinone molecule) diffuses inside the membrane, between the sites A and H (on the P-side) and the sites B and
L (on the N-side). As its biological counterpart, the shuttle has two electron sites and two proton-binding sites. At its N-position, the
shuttle takes one electron from the site A and another electron from the site H and transfers these electrons to the sites B and L.

At the N-side, the shuttle also accepts up to two protons from the stromal (electrically-negative) proton reservoir and donate these protons to
the lumenal proton reservoir at the P-side of the membrane. When the fully populated Q-molecule arrives at the P-side, one electron from the
shuttle goes strictly energetically downhill, to the site B, whereas another one returns to the  L-H chain to be loaded again on the shuttle.
The origin of this bifurcated reaction \cite{Wikstrom72}, which occurs at the P-side catalytic center, remains unknown.

Here, we explore physicochemical conditions wherein our artificial complex is able to translocate twice as protons
 as the number of electrons transferred energetically downhill, from the source S to the drain D. Protons are translocated energetically uphill,
 from the N- to P-side of the membrane. We aim at the explanation of the Q-cycle operation in this artificial complex. We examine
a wide range of parameters allowing the efficient performance of the Q-cycle scheme. The functional principles
  of the Q-cycle in artificial systems can provide a better understanding of the Q-cycle mechanism in the natural $bc_1$ and $bf$ complexes.

\section{Model and methods}
\subsection{Components and states}
To simplify the problem, we divide the whole system of six electron and two proton-binding sites into four weakly-interacting subsystems: (i)
the LH subsystem consisting of L and H sites (with energies $\varepsilon_L, \varepsilon_H$); (ii) the shuttle Q having two electron sites
$1_e,2_e $ (with energies $\varepsilon_1 = \varepsilon_2 = \varepsilon_Q$) and two proton sites $1_p, 2_p$ (with energies $E_1 = E_2 = E_Q$),
(iii) the site A (with energy $\varepsilon_A$), and (iv) the site B (with energy $\varepsilon_B$) .

The LH-subsystem is characterized by 4 microscopic states starting with the empty (vacuum) state, $|1\rangle_{LH} = |0_L 0_H\rangle,$ and ending
with the doubly-occupied system: $|4\rangle_{LH} = |1_L 1_H\rangle.$ The electron and proton populations of the shuttle are described by 16
states, where $|1\rangle_{Q} = |0_{1e} 0_{2e} 0_{1p} 0_{2p}\rangle$ is the vacuum state and $|16\rangle_{Q} = |1_{1e} 1_{2e} 1_{1p}
1_{2p}\rangle$ is the state of the completely-loaded shuttle. Here, we use the notation $0_{\alpha}\,(1_{\alpha})$ for an empty (occupied) site
$\alpha$, where $\alpha = 1_e,\,2_e$ for the electron sites on the shuttle, and $\alpha = 1_p,\,2_p$ for the proton sites. The average
populations of the sites A and B are denoted by $\langle n_A\rangle $ and $\langle n_B\rangle$. Here $\langle \ldots \rangle = \langle \langle
\Psi_0|\ldots |\Psi_0 \rangle \rangle_T$ means double-averaging over an initial wave function $\Psi_0$ and over a thermal distribution $\langle
\ldots \rangle$ of reservoirs and environment characterized by the common temperature $T$.

 We take into account strong Coulomb
interactions between sites from the same subsystem. A Coulomb repulsion between electrons located on the sites L and H (LH-subsystem) is
characterized by the energy $u_{LH}$, whereas for the electrons and protons on the shuttle (Q-subsystem) we introduce the following parameters:
$U_e$ (Coulomb repulsion between two electrons occupying the shuttle sites $1_e,2_e$); $U_p$ (electrostatic repulsion between two protons
located on the sites $1_p, 2_p$); and $U_{ij} = U_{ep}$ (electron-proton Coulomb attraction on the shuttle). Here the indices $i=1,2$ and
$j=1,2$ run over the electron and proton-binding sites of the shuttle, respectively.

\subsection{Electron and proton transitions}

 Electrons can tunnel between the L and H sites as well as between sites belonging to different subsystems.
 These tunnelings are described by relatively small tunneling amplitudes: $\Delta_{LH}$ (L-H tunneling); $\Delta_{HQ} (x)$
(tunneling between the H site and the $1_e,2_e$ sites; $\Delta_{LQ} (x)$ (for electron transitions between the $1_e,2_e$ and L sites. The
amplitudes $\Delta_{AQ} (x)$  and $\Delta_{BQ} (x)$ describe the tunneling between the bridge sites A, B and the electron sites $1_e,2_e$ on the
shuttle. The amplitudes $\Delta_{AQ}, \Delta_{HQ}$ and $\Delta_{BQ}, \Delta_{LQ}$ depend on the position  $x$ of the mobile molecule Q. The
proton rates $\Gamma_N(x)$ (for transitions between the sites $1_p,2_p$ and the N-side proton reservoir) and $\Gamma_P(x)$ (for transitions
between the proton sites on the shuttle and the P-reservoir) are also functions of $x$.

The time evolution of $x(t)$ is determined by an overdamped Langevin equation with terms describing walls of the membrane and a potential
barrier, which impedes the charged shuttle to cross the intermembrane space \cite{SmirnovPRE09}. The walls of the membrane are located at $x = -
x_0$ (N-side) and at $x = x_0$ (P-side), so that the width of the membrane is equal to $2x_0$.

\subsection{Equations and potentials}

 Using the formalism outlined in Ref.~\cite{SmirnovPRE09,GhoshJCP09,SmirnovJPC09} and based on Marcus transitions rates~\cite{Krish01}, we
derive and numerically solve a set of master equations (see an Appendix) for an electron distribution $\langle R_M\rangle $ over states of the
LH-subsystem coupled to a system of equations for the probabilities $\langle \rho_{\mu}\rangle $ to find the Q-subsystem in the state
$|\mu\rangle$. These equations are complemented by equations for the populations $\langle n_A\rangle,\, \langle n_B\rangle$ of the A and B-sites
and by an overdamped Langevin equation for the shuttle position $x(t)$.

It is known \cite{Nicholls02} that charged ions, such as Mg$^{2+}$ and Cl$^{-}$, can easily cross a thylakoid membrane, which results in the
equilibration of electrical potentials on both sides of a membrane. Therefore, the difference between the proton electrochemical potential
$\mu_P$ of the P-side, and the potential $\mu_N$ of the N-side of the membrane \cite{Nicholls02} is mainly determined by the proton
concentration gradient $\Delta pH$ :
\begin{equation}
\mu_P - \mu_N \simeq -  \Delta pH \times (T/T_R)\times 60~{\rm meV}, \label{muPN}
\end{equation}
 where $T$ is the temperature of the reservoirs, and $T_R = 298$~K is the room temperature. It should be noted, however, that
a surface potential $V_S(x)$, which is positive on the N-side, $V_S(-x_0) = + V_N$, and negative, $V_S(x_0) = -V_P$, on the P-side of the
membrane, was calculated for the $bf$ complex of \emph{M. laminosus} \cite{CramerAR06} with $V_N = 4.6 \,T $ and $V_P = 5.4 \,T $ (the Boltzmann
constant $k_B = 1$). This model assumes that there is a similar transmembrane potential (see Fig.~1),
\begin{equation}
V_S(x) =  - \frac{x-x_0}{2 x_0} \,V_N -  \frac{x+x_0}{2 x_0}\, V_P,
\end{equation}
with $V_N \simeq 120$~meV and $V_P \simeq 140$~meV, which correspond to the above-mentioned values of $V_N$ and $V_P$ at room temperature. All
energies are measured in meV. In the presence of the surface potential, the energy levels of electrons and protons on the shuttle are shifted
from their initial values $\varepsilon_{Q0}$ and $E_{Q0} $ depending on the shuttle's position $x$:
\begin{eqnarray}
\varepsilon_Q(x) = \varepsilon_{Q0} - V_S(x), \nonumber\\
E_Q(x) = E_{Q0} + V_S(x),
\end{eqnarray}
with $\varepsilon_{QN} = \varepsilon_{Q0} - V_N, \; \varepsilon_{QP} = \varepsilon_{Q0} + V_P$ and $E_{QN} = E_{Q0} + V_N, \; E_{QP} = E_{Q0} -
V_P.$
 Correspondingly, the energy levels of the electron sites A and H, located near the N-side, are shifted down from their initial values:
$\varepsilon_A = \varepsilon_{A0} - V_N, \; \varepsilon_H = \varepsilon_{H0} - V_N $; whereas the energies of the sites B and L, located near
the P-side, are shifted up: $\varepsilon_B = \varepsilon_{B0} + V_P, \; \varepsilon_L = \varepsilon_{L0} + V_P. $

\section{ ``Passenger" scenario of the Q-cycle }

Instead of searching over a multidimensional space of system parameters, we consider a reasonable sequence of events, which provides an optimal
performance of the energy transducer. In particular, we analyze a scenario where an electron transferred from the high-energy source reservoir
$S$ along the chain: $ S \rightarrow A \rightarrow Q \rightarrow B \rightarrow D $ to the drain $D$ performs the main energetic function in the
transfer of two protons from the N to the P-side of the membrane. Another electron traveling on the shuttle and recycled by the LH-system along
the chain $H \rightarrow Q \rightarrow L \rightarrow H$ plays a more passive role of a passenger, which is necessary to compensate a shuttle
charge.

 According to the Marcus formula \cite{Krish01} (see also an Appendix),
\begin{equation}
\kappa_{ii'} = |\Delta_{ii'}|^2 \sqrt{\frac{\pi}{\lambda_{ii'} T}}\; \exp\left[ - \frac{(\varepsilon_{i} - \varepsilon_{i'} - \lambda_{i i'})^2
}{4 \lambda_{ii'} T} \right], \label{MarcusRate}
\end{equation}
the rate $\kappa_{ii'}$ for an electron transition from the site $i$, with an energy $\varepsilon_i$, to the site $i'$, with an energy
$\varepsilon_{i'}$, has a maximum at $\varepsilon_{i} = \varepsilon_{i'} + \lambda_{ii'}.$ Here $\Delta_{ii'}$ is the tunneling amplitude
between the sites $i$ and $i'$, $\lambda_{ii'}$ is the corresponding reorganization energy, which is due to electron coupling to an environment
with temperature $T$. The shuttle can accept protons from the N-reservoir provided that the electrochemical potential of the N-side, $\mu_N$, is
higher than the proton energy level on the shuttle. Protons move from the shuttle Q to the P-side reservoir if the energy of the Q-proton
exceeds the P-side potential $\mu_P$.

\subsection{Sequence of events and energy relations}
 We start with a situation when the empty shuttle (quinone) is near the N-side catalytic center, $x = - x_0$, and  the $LH$-system
 (analog of cytochrome $b$ in the $bf$ complex) is preloaded with one electron located
presumably at the site $H$, which has a lower energy than the site $L$:   $\varepsilon_L > \varepsilon_H $. The site A is also occupied with an
electron taken from the electron source S. We have the following sequence of electron ($e$) and proton ($p$) transfer from and to the shuttle
located near the N-side of the membrane: \vspace{0.25cm}
 \newline a) $e :\,H\rightarrow Q, \; \; \varepsilon_H = \varepsilon_{QN} + \lambda_{HQ}$.
 \newline b) $e :\,A\rightarrow Q, \; \; \varepsilon_A = \varepsilon_{QN} + U_e +  \lambda_{AQ}$.
\newline c) $p :\; N\rightarrow Q, \;\; \mu_N > E_{QN} - 2 \, U_{ep}$.
\newline d) $p :\; N\rightarrow Q, \;\; \mu_N > E_{QN} - 2 \, U_{ep} + U_p$.
\vspace{0.25cm}
\newline
 Here we have written relations between energy levels of electrons and protons which make possible these transfers.

The shuttle loaded with two electrons and two protons travels to the P-side of the membrane, where the following sequence of electron and proton
transitions occurs: \vspace{0.25cm}
\newline e) $e :\, Q\rightarrow B, \; \; \varepsilon_{QP} + U_e - 2 U_{ep} = \varepsilon_B + \lambda_{BQ}$.
\newline f) $p :\; Q\rightarrow P, \;\;  E_{QP} - U_{ep} + U_p > \mu_P$.
\newline g) $p :\; Q\rightarrow P, \; \; E_{QP} - U_{ep} > \mu_P$.
\newline h) $e :\, Q\rightarrow L, \; \;  \varepsilon_{QP} = \varepsilon_L + \lambda_{LQ}$. \vspace{0.25cm}
\newline Finally, an electron tunnels from the L  to the H site: \vspace{0.25cm}
\newline i) $e :\, L\rightarrow H, \; \; \varepsilon_{L} = \varepsilon_H + \lambda_{LH}$. \vspace{0.25cm}
\newline The empty shuttle diffuses to the N-side of the membrane and the process repeats. We expect that two protons will be translocated from
the N-side to the P-side of the membrane per one electron transferred from the source to the drain electron reservoir with a quantum yield $QY =
2.$

Here we assume that, as in the case of the quinone molecule Q \cite{OkamuraBBA00}, the shuttle populated with one electron (after step 1) does
not bind a proton but accepts another electron (step 2). The doubly-reduced quinol is known to have a much stronger ability for binding two
protons (see steps 3 and 4). At the P-side of the membrane the process presumably evolves in the opposite direction when the transfer of one
electron from Q to the site B is accompanied by the unloading of two protons. In the absence of an attraction to two positive charges, the
energy of the electron remaining on the shuttle goes up; thus, allowing its tunneling to the L-site.

Here, an electron recycled by the LH-system plays a passive role of a shuttle's ``passenger" since its transitions to and from the Q-molecule
are not immediately accompanied by a proton transfer. Transitions of another electron, which is loaded to the shuttle from the source S (via the
site A) and unloaded to the drain D (via the site B), are more closely coupled to the energetically-uphill proton translocation.

It follows, from the relations a),\,h),\,i) in this section, that the recycling of one electron by the LH-chain, H $\rightarrow$ Q $\rightarrow$
L $\rightarrow$ H, which lies at the heart of the Q-cycle, takes place if the difference of surface potentials,
\begin{equation}
\Delta V = V_S(-x_0)- V_S(x_0) = V_N + V_P,
\end{equation}
 is of the order of the total reorganization energy along the recycling path:
\begin{equation} \label{deltaV}
\Delta V = \lambda_{HQ} + \lambda_{LQ} + \lambda_{LH}.
\end{equation}
We see from the relations d) and g) in this section that the energetically-uphill proton transfer from N- to the P-side of the membrane is
possible if the original energy of the proton on the shuttle, $E_{Q0}$, obeys the following inequality:
\begin{equation} \label{EQ0}
\mu_{N} + 2\, U_{ep} - U_p  - V_N \, > \,  E_{Q0} \, > \, \mu_{P} +U_{ep} + V_P,
\end{equation}
which can be true only for a sufficiently strong attraction potential, $U_{ep},$ between electrons and protons on the shuttle
\begin{equation} \label{Uep}
U_{ep} \, > \, \mu_P - \mu_N + \Delta V + U_p.
\end{equation}

 The relations  a),\,b),\,e),\,h),\,i) in this section allow to estimate the original energies of the electron-binding sites counted,
e.g,, from the level
$\varepsilon_{B0}$:
\begin{eqnarray} \label{energy0}
\varepsilon_{A0} &=& \varepsilon_{B0} + 2 U_{ep} + \lambda_{AQ} + \lambda_{BQ}, \nonumber\\
\varepsilon_{H0} &=& \varepsilon_{B0} + 2 U_{ep} - U_e + \lambda_{BQ} + \lambda_{HQ}, \nonumber\\
\varepsilon_{Q0} &=& \varepsilon_{B0} + 2 U_{ep} -U_e + \lambda_{BQ}, \nonumber\\
\varepsilon_{L0} &=& \varepsilon_{B0} + 2 U_{ep} - U_e  + \lambda_{BQ} - \lambda_{LQ}.
\end{eqnarray}
We assume that the potentials of the electron source, $\mu_S$, and electron drain, $\mu_D$, are of the order of the energies of the A and B
sites, respectively: $\mu_S = \varepsilon_A, \, \mu_D = \varepsilon_B$. Taking into account Eqs.~(\ref{energy0}) we obtain a relation for the
source-drain energy drop,
\begin{equation}
\mu_S - \mu_D \, \geq \, 2 \,U_{ep} + \lambda_{AQ} + \lambda_{BQ} - \Delta V.
\end{equation}
With Eqs.~(\ref{deltaV},\ref{Uep}) we obtain the following requirement for the energy difference between the source and drain electron
reservoirs:
\begin{equation} \label{muSD}
\mu_S - \mu_D \, > \, 2 \,(\mu_P - \mu_N) + 2\, U_p + \lambda_{\rm tot},
\end{equation}
where the combined reorganization energy,
\begin{equation}
\lambda_{\rm tot} = \lambda_{AQ} + \lambda_{BQ} + \lambda_{HQ} + \lambda_{LQ} + \lambda_{LH},
\end{equation}
accumulates all losses along both electron transport chains: $ A \rightarrow Q \rightarrow B$ and $ H \rightarrow Q \rightarrow L \rightarrow
H$.

\subsection{Thermodynamic efficiency and quantum yield}

The thermodynamic efficiency $\eta$ of proton translocation  can be defined as
\begin{equation}
\eta = \frac{\mu_P - \mu_N}{\mu_S - \mu_D} \times \frac {N_P}{n_D},
\end{equation}
where $N_P$ is the number of protons translocated from the N- to the P-side of the membrane, and $n_D$ is the number of electrons transferred
from the source S to the electron drain D. The efficiency $\eta$ is proportional to the quantum yield
\begin{equation}
QY = \frac{N_P}{n_D}.
\end{equation}
 It follows from Eq.~(\ref{muSD}) that, within the ``passenger" scenario, the efficiency $\eta $ of the electron-to-proton energy conversion
can be estimated as
\begin{equation} \label{EqEta}
\eta = \frac{\mu_P - \mu_N} { 2\,(\mu_P - \mu_N) + 2\, U_p + \lambda_{\rm tot} } \times QY.
\end{equation}
This means that for a high electrochemical proton gradient, $ \mu_P - \mu_N \gg U_p + 0.5\,\lambda_{\rm tot},$ the efficiency $\eta$ has the
maximum: $\eta = QY/2.$ Thus, in the ideal case, when $QY = 2$, the thermodynamic efficiency can reach the perfect mark: $\eta =1,$ when almost
all electron energy is converted to the transmembrane proton-motive force.

\section{Results and discussions}

\subsection{Parameters}

In the model presented here an electron transport chain begins at the source reservoir S, corresponding to a pool of ferredoxin (Fd) molecules,
which carry electrons from Photosystem~I to the $bf$ complex (see Fig.~1). The electron drain D is related to the high-potential chain of the
$bf$ complex comprised of the ion-sulfur protein (ISP), cytochrome $f$, and soluble plastocyanin (PC) molecules. Taking into account  a redox
potential of ferredoxin, $E_m = - 0.41$~V, and the fact that a redox potential of the ISP/$f$/PC chain, $E_m$ is in the range from 0.3 to 0.45~V
\cite{CramerAR06}, we estimate that the total energy drop between the source and the drain, $\mu_S - \mu_D$, takes values from 710 meV up to 860
meV.

We also assume that, as for the $bf$ complex \cite{CramerAR06}, the surface voltage gradient, $\Delta V = V_N + V_P = 260$~meV, has been applied
to the membrane, with a positive potential, $V_S(-x_0) = V_N = 120$~meV, at the N-side and a negative potential, $V_S(x_0) = -V_P = - 140$~meV,
at the P-side of the membrane.

The system reaches its optimal performance when $U_{ep} = 610$~meV. Hereafter, we assume that $U_e = U_{ep}/2, \,U_p = U_{ep}/8, \, U_{LH} =
240$~meV. We use the following values for the electron transfer rates, $\gamma_S = \gamma_D = 0.1~\mu$eV, proton transition rates, $\Gamma_N =
\Gamma_P = 2~\mu$eV, and for peak values of the electron tunneling amplitudes, $\Delta_{AQ}(-x_0) = \Delta_{DQ}(x_0) = 0.1$~meV,
$\Delta_{HQ}(-x_0) = \Delta_{LQ}(x_0) = 0.06$~meV. The coefficients $\gamma_S $ and $\gamma_D$ determine the rates of the electron transfer
between site A and source S and between site B and drain D, respectively (see the definitions of $\gamma_S$ and $\gamma_D$ in the Appendix). The
rates $\Gamma_N$ and $\Gamma_P$, describing proton transitions between the N-side reservoir and the shuttle as well as between the shuttle and
the P-side of the membrane, respectively,  are defined in the Appendix and in the subsection IE of the Supporting Information. For the diffusion
coefficient, ${\cal D} = T/\zeta$, we have the value, ${\cal D} \simeq 8 \times 10^{-12}$~m$^2$/s, which is close to experimental data for
plastoquinone molecules in a lipid membrane \cite{Marchal98}. Hereafter, we assume that the reorganization energies, corresponding to H-Q and
L-Q transitions, are equal: $\lambda_{HQ} = \lambda_{LQ}$, and the same relation is true for the A-Q and B-Q transitions: $\lambda_{AQ} =
\lambda_{BQ}.$

To satisfy Eq.~(\ref{deltaV}) we start with small values of the reorganization energies along the recycling pathway, $\lambda_{LQ} =
\lambda_{HQ} = 60$~meV, and $\lambda_{LH} = 140$~meV, which add up to the value of the surface gradient, $\Delta V = 260$~meV. Here, we also
have $\lambda_{AQ} = 60$~meV. For the driving force, $\mu_S-\mu_D = 850$~meV, related to the electron energy drop in the $bf$ complex, the
system is able to translocate almost two protons, $QY \simeq 1.9$, against the electrochemical gradient, $\mu_P - \mu_N = 150$~meV, per each
electron transferred from the source to the drain. This proton electrochemical gradient corresponds to the value $\Delta pH = -2.5$ at room
temperature, $T = T_R = 298$~K. The thermodynamic efficiency of the energetically-uphill proton translocation is about 33 \% ($\eta \simeq
0.33$).

\subsection{Time evolution of a proton translocation process}

A proton translocation process is shown in Fig.~2, where we plot the time dependence of the total electron, $n_Q = \langle n_1\rangle  + \langle
n_2\rangle$, and proton, $N_Q = \langle N_1\rangle  + \langle N_2\rangle$, populations of the shuttle (Fig.~2b), together with the time-evolving
position of the shuttle $x(t)$ (Fig.~2a). Here, we also show (see Fig.~2c) the populations of the L-site, $\langle n_L\rangle$, and H-site,
$\langle n_H\rangle$, as well as the average number of electrons $\langle n_D\rangle$ transferred from the source to the drain, as well as the
average number of protons $\langle N_P\rangle$ translocated from the N-side to the P-side proton reservoir. The brackets $\langle \ldots\rangle$
are dropped in Figs.~2c and 2d for the notations of the populations, and throughout the paper, except in the Appendix. Data for Fig.~2 are
calculated at higher values of the reorganization energies: $\lambda_{LQ} = \lambda_{HQ} = 100$~meV, $\lambda_{LH} = 250$~meV, and $\lambda_{AQ}
= 100$~meV. We assume that $U_{ep} = 610$~meV, $\Delta V = 260$~meV, $\mu_S-\mu_D = 850$~meV, and $\mu_P - \mu_N = 150$~meV.

It can be seen from Fig.~2 that during 30~$\mu$s the shuttle performs about 8 trips from the N-side ($x = - x_0$) to the P-side ($x = x_0 =
2$~nm) of the membrane and back, translocating in the process about 7 electrons ($ n_D $ = 6.7) and 12 protons ($ N_P = 12.2$), with quantum
yield $QY \simeq 1.8$, and thermodynamic efficiency $\eta = 32 \%$.

At the N-side (see Fig.~2b) the shuttle accepts an electron from the initially populated site H and another electron from the source S (via site
A) as well as two protons from the N-side proton reservoir. We note (see Fig.~2c) that site H is not completely depopulated. This means that
there are events when both electrons occupying the shuttle arrive from the site A and the source S, shorting out the  Q-cycle pathway. This
leakage process increases the number of electrons transferred from the source to the drain (with the same number of protons), thus decreasing
the quantum yield $QY$.

Here we do not impose any additional restrictions, except a proper choice of energy levels, which are close to values given by
Eqs.~(\ref{EQ0},\ref{energy0},\ref{muSD}), with $ \mu_S = 410,\,\varepsilon_A = 465,\, \varepsilon_H = 220, $ and $\varepsilon_{QN} = 160,\,
E_{QN} = 982,$ at the N-side catalytic center (all energies are measured in meV). For the P-side center we use the following energies:
$\varepsilon_{QP} = 420,\,E_{QP} = 722,$ for electrons and protons on the shuttle,  $ \varepsilon_B = -495,\, \mu_D = -440,$ for the
high-potential redox chain, and $\varepsilon_L = 360$~meV for the recycling pathway. This choice of energy levels makes the H-to-Q electron
transition (at the N-side) much easier than the A-to-Q electron transfer, since $\varepsilon_H - \varepsilon_{QN} \leq \lambda_{HQ}$, whereas
$\varepsilon_A - \varepsilon_{QN} \gg \lambda_{AQ}$. Moreover, the S-to-A electron transition is also hampered since the energy level of the
A-site, $\varepsilon_A$, is higher than the potential of the source, $\mu_S.$

On arrival at the P-side of the membrane the shuttle donates an electron to the B-site and, finally, to the drain D. Two protons move to the
P-side proton reservoir (Figs.~2b and 2d).  It is evident from Fig.~2c that another electron from the shuttle Q goes to the L-site (see the
small spike at the bottom of Fig.~2c). This electron is rapidly transferred to the H-site, and the empty shuttle returns to the N-side.

Figures 2b,\,2c,\,2d illustrate the bifurcated reaction which occurs at the P-side. Here, one electron from the shuttle Q goes to the
high-potential (and low-energy) chain, $Q \rightarrow B \rightarrow D,$ while another electron (a passenger) returns to the LH-system for
recycling (along the pathway $Q \rightarrow L \rightarrow H$). No additional gate mechanisms are required for this reaction. An escape of the
first electron from Q to B, followed by the transition of two protons to the P-reservoir, increases the energy of the remaining electron to the
level $\varepsilon_{QP} = 420$~meV, which is of order of the L-site energy, $\varepsilon_L = 360$~meV, but is much higher than the energy of the
B-site, $\varepsilon_B = -495$~meV. Furthermore, the site B is probably occupied with an electron taken from the drain since $\mu_D >
\varepsilon_B.$ These two factors strongly suppress the leakage of the second electron from the shuttle to the high-potential chain.

\subsection{Effects of the proton electrochemical gradient}

In Figure 3 we show the numbers of protons, $N_P$, and electrons, $n_D$, transferred across the membrane, as well as the quantum yield, $QY$,
and the power-conversion efficiency, $\eta$, as functions of the proton electrochemical gradient, $\mu_P - \mu_N$ (measured in meV). The graphs
are plotted for two sets of reorganization energies: (i) $\lambda_{LQ}=\lambda_{HQ} = 100$~meV, $\lambda_{LH} = 250$~meV,
$\lambda_{AQ}=\lambda_{BQ} = 100$~meV (blue curves) (ii) $\lambda_{LQ}=\lambda_{HQ} = 200$~meV, $\lambda_{LH} = 400$~meV,
$\lambda_{AQ}=\lambda_{BQ} = 200$~meV (green dashed curves). Other parameters, such as $U_{ep} = 610,\,\mu_S~-~\mu_D = 850,\, \Delta V = 260$
(in meV), are the same as in Fig.~2.

We numerically calculate the output of the system ($N_P,\,n_D,$ etc.) at the end of the stochastic trajectory $x(t)$ (with the duration
$t=100$~$\mu$s) and average results over 10 trajectories. For each value of the shuttle's position $x(t)$ we solve a set of master equations
(\ref{RMN},\ref{rhoMuNu},\ref{nA}), which have been averaged over electron and proton reservoirs as well as over fluctuations of the environment
coupled to the electronic degrees of freedom. This can be done since the electron and proton transitions are much faster than the mechanical
motion of the shuttle.

It follows from Fig.~3 that more protons, $N_P \simeq 45$, and electrons, $n_D \simeq 25$, are transferred across the membrane at lower proton
gradients, $\mu_P - \mu_N \leq 150$~meV with a higher quantum yield, $QY \geq 1.8$. However, the thermodynamic efficiency is higher, $\eta
\simeq 39 \%$,  at larger proton gradients, $\mu_N - \mu_P \simeq 200$~meV, where $N_P \sim 40$ and $QY \sim 1.6$. These numbers are for the set
(i), with smaller values of the reorganization energies (see blue continuous curves in Fig.~3). A stronger electron-environment interaction,
described by the set (ii) of reorganization energies, significantly reduces an energetically-uphill proton flow with almost no impact on the
electron current (see green dashed curves in Fig.~3). In this case the quantum yield drops to almost one, which means that the recycling pathway
(via the LH-system) is practically closed.

\subsection{Effects of the surface potential gradient}

Figure 4 demonstrates the performance of the system as a function of the surface potential gradient, $\Delta V = V_N + V_P,$ at a fixed
difference between the N-side and P-side potentials, $V_P - V_N = 20$~meV. Here we choose two sets of system parameters. The first set (see the
blue continuous curves in Fig.~4), with $U_{ep} = 610, \, \mu_S - \mu_D = 850,$ and $\mu_P - \mu_N =150$~meV, was considered before. The second
set (green dashed curves in Fig.~4) is characterized by a higher electron-proton attraction potential, $U_{ep} = 800$~meV, and a higher
source-drain difference, $\mu_S - \mu_D = 1220$~meV. At these parameters the system can translocate protons against the electrochemical
difference $\mu_P - \mu_N = 300$~meV, which is related to the proton concentration gradient $\Delta pH = - 5$ at room temperature. The
reorganization energies correspond to the set (i) described before.

It follows from Fig.~4 that the system performs very well in a wide range of surface voltage gradients, from $\Delta V \sim 230$~meV up to
$\Delta V \sim 350$~meV, translocating more than 40 protons and 20 electrons (in 100 $\mu$s). These numbers are almost the same for both sets of
parameters (see Figs.~4a and 4b).  The quantum yield $QY$ monotonically goes down, from $QY \sim 2$ at $\Delta V \sim 200$~meV to $QY \sim 1.5$
at $\Delta V \sim 350$~meV, no matter what set is chosen. The power-conversion efficiency $\eta$ also decreases with increasing $\Delta V$.
However, $\eta $ is higher for the second set of parameters. For example, at $\Delta V = 260$~meV, the efficiency is about 31\%  for the first
set, and $\eta \sim 47 \%$ for the second set (with $QY = 1.9$) where both electron and proton electrochemical gradients are higher. This trend
is consistent with Eq.~(\ref{EqEta}) derived for the passenger scenario.

\section{Conclusions}
We have theoretically examined a model of the protonmotive force generation by the Q-cycle mechanism mimicking the operation of the $bf$ complex
in the thylakoid membranes of plants and cyanobacteria. We concentrate on a simulation of the regime of ferredoxin-dependent cyclic electron
flow, where the $bf$ complex translocates up to two protons (a quantum yield $QY=2$) across the membrane per one electron transferred from the
electron source (ferredoxin pool) to the electron drain (pool of plastocyanin molecules). This model includes two electron and two proton sites
on the shuttle Q (an analog of a plastoquinone molecule), diffusing inside the membrane, as well as two electron sites, A and B, connecting the
electron sites on the shuttle to the source and drain reservoirs. The recycling of an electron by the cytochrome $b$, which forms the basis of
the Q-cycle, can be described by adding two electron sites L and H corresponding to hemes $b_L$ and $b_H$ of the $bf$ complex.

We have derived and numerically solved a set of master equations for the populations of the electron and proton-binding sites together with a
Langevin equation for the position of the shuttle. Within a reasonable scenario and in the presence of the surface potential, we have determined
the conditions which are necessary for the efficient translocation of protons across the membrane. We have found  that the system is able to
transfer, on average, about 1.8 protons per one electron ($QY = 1.8$) with a thermodynamic efficiency of the order of 32\% against the
transmembrane proton gradient $\mu_P - \mu_N = 150$~meV at the source-drain difference of electron potentials $\mu_S - \mu_D = 850$~meV. These
values of the electron and proton gradients are closely related to experimental values for the $bf$ complex. No conformational gating is
necessary for the bifurcation of the electron transfer reaction at the P-side catalytic center, where one electron goes to the drain and another
electron returns to the LH-chain, to be loaded on the shuttle again. We have studied the performance of the model as a function of the proton
electrochemical gradient and the surface potential. It is shown that the system demonstrates even better results, with a quantum yield of the
order of 1.9 and an efficiency of the order of 47\%, when both the source-drain difference and the proton gradient are higher.

\textbf{Acknowledgements.} FN acknowledges partial  support from the Laboratory of Physical Sciences, National Security Agency, Army Research
Office, DARPA, Air Force Office of Scientific Research, National Science Foundation grant No. 0726909, JSPS-RFBR contract No. 09-02-92114,
Grant-in-Aid for Scientific Research (S), MEXT Kakenhi on Quantum Cybernetics, and JSPS via its FIRST program.

\begin{appendix}
\section{Master equations for electron-driven proton transfer across a membrane}
Here  we briefly outline a set of master equations describing the process of electron-driven proton translocation across a membrane. As we
mentioned before, the total system is characterized by the average populations of the A and B-sites, $\langle n_A\rangle,\, \langle n_B
\rangle$, as well as by four states of the LH-subsystem with electron distributions $\langle R_M\rangle $ ($M = 1,\ldots,4$) and 16 states of
the Q-subsystem (electrons and protons on the shuttle) with distributions $\langle \rho_{\mu}\rangle $ ($\mu = 1,\ldots,16$).

The time evolution of the LH-system is governed by the equation
\begin{equation} \label{RMN}
\langle \dot{R}_M\rangle =  - \sum_N \gamma_{NM}^{LH} \; \langle R_M\rangle + \sum_N \gamma_{MN}^{LH}\; \langle R_N\rangle,
\end{equation}
with the following relaxation matrix
\begin{equation}
 \gamma_{MN}^{LH} = \gamma_{MN}^{\rm tun} + \gamma_{MN}^{LQ} +  \gamma_{MN}^{HQ}.
 \end{equation}
Here the rate
\begin{eqnarray}
\gamma_{MN}^{\rm tun} = |\Delta_{LH}|^2 \sqrt{\frac{\pi}{\lambda_{LH}T}}\times  \{ \, |\langle M|a_L^\dagger a_H|N\rangle |^2 + |\langle
N|a_L^\dagger a_H|M\rangle |^2 \, \}\nonumber\\ \times  \exp \left[\, - \,\frac{(\Omega_{MN} + \lambda_{LH})^2}{4\lambda_{LH}T}\,\right]
\end{eqnarray}
describes the L-to-H electron transitions. Hereafter, $a_{\alpha}^\dag, a_{\alpha}$ refer to the creation/annihilation operators for an electron
on the site $\alpha$. For protons on the site $\beta$ the creation/annihilation operators are denoted by $A_{\beta}^\dag, A_{\beta}.$ The rate
$\gamma_{MN}^{LQ}$ (and a similar rate $\gamma_{MN}^{HQ}$) is related to the electron transfer between the L (or H) sites and the sites
$1_e,\,2_e$ on the shuttle,
\begin{eqnarray}
\gamma_{MN}^{LQ} = |\Delta_{LQ}|^2 \sqrt{\frac{\pi}{\lambda_{LQ}T}}\, \sum_{\mu\nu}\, |\langle \mu |a_1+a_2|\nu\rangle|^2 \times \nonumber\\
\left\{\, |\langle M|a_L|N\rangle |^2\, \exp \left[\, - \, \frac{(\Omega_{MN} - \omega_{\mu\nu} + \lambda_{LQ})^2}{4\lambda_{LH}T}\, \right]\,
\langle \rho_{\mu}\rangle + \right. \nonumber\\ \left. |\langle N|a_L|M\rangle |^2\, \exp \left[\, -\, \frac{(\Omega_{MN} + \omega_{\mu\nu} +
\lambda_{LQ})^2}{4\lambda_{LH}T}\,\right] \, \langle \rho_{\nu}\rangle \right\},
\end{eqnarray}
where $a_1,a_2$ are operators of the electron-binding sites on the shuttle, $\Omega_{MN} = E_M - E_N$ are frequencies of the LH-system ($\hbar =
1,\, k_B = 1$), and $\omega_{\mu\nu}$ is the frequency spectrum of the coupled electron-proton states on the shuttle Q.

For the distributions $\langle \rho_{\mu}\rangle $ of the 16 states of the Q-system we derive the following equation
\begin{equation} \label{rhoMuNu}
\langle \dot{\rho}_{\mu} \rangle =  - \sum_{\nu}  \gamma_{\nu\mu}^{Q}\, \langle \rho_{\mu}\rangle  + \sum_{\nu}  \gamma_{\mu\nu}^{Q} \,
 \langle \rho_{\nu}\rangle,
 \end{equation}
where
 \begin{equation}
 \gamma^Q_{\mu\nu} = \gamma^{AQ}_{\mu\nu} + \gamma^{BQ}_{\mu\nu} + \gamma^{LQ}_{\mu\nu} + \gamma^{HQ}_{\mu\nu} + \gamma^{NQ}_{\mu\nu}
 + \gamma^{PQ}_{\mu\nu}.
 \end{equation}
Components of this relaxation matrix can be written as
\begin{eqnarray}
\gamma_{\mu\nu}^{AQ} = |\Delta_{AQ}|^2 \, \sqrt{\frac{\pi}{\lambda_{AQ} T}} \times \nonumber\\
 \left\{\, |\langle \mu|a_1+a_2|\nu\rangle |^2 \, \exp\left[\, - \, \frac{
(\omega_{\mu\nu} + \varepsilon_A + \lambda_{AQ})^2}{4 \lambda_{AQ}T} \, \right] \langle 1 - n_A\rangle
 \right. \nonumber\\ \left. + |\langle \nu|a_1+a_2|\mu\rangle |^2 \, \exp\left[\, -\,
\frac{ (\omega_{\mu\nu} - \varepsilon_A + \lambda_{AQ})^2}{4 \lambda_{AQ}T} \, \right] \langle n_A\rangle \right\},
\end{eqnarray}
with a similar matrix $\gamma_{\mu\nu}^{BQ}$, and
\begin{eqnarray}
\gamma_{\mu\nu}^{LQ} = |\Delta_{LQ}|^2 \, \sqrt{\frac{\pi}{\lambda_{LQ} T}} \, \sum_{MN}\, |\langle M |a_L| N\rangle|^2 \times \nonumber\\
\left\{\, |\langle \nu|a_1+a_2|\mu\rangle |^2 \, \langle R_N\rangle \, \exp\left[\, - \, \frac{ (\omega_{\mu\nu} + \Omega_{MN} +
\lambda_{LQ})^2}{4 \lambda_{LQ}T} \, \right] \right. \nonumber\\ \left.
 + \, |\langle \mu|a_1+a_2|\nu\rangle |^2 \, \langle R_M\rangle \, \exp\left[\, - \, \frac{ (\omega_{\mu\nu} - \Omega_{MN} +
 \lambda_{AQ})^2}{4 \lambda_{LQ}T} \, \right]\,  \right\},
\end{eqnarray}
with a matrix $\gamma_{\mu\nu}^{HQ}$, which is similar to $\gamma_{\mu\nu}^{LQ}$. The proton transitions to and from the shuttle are described
by the rate
\begin{eqnarray}
\gamma_{\mu\nu}^{NQ} = \Gamma_N |\langle \mu|A_1 + A_2|\nu\rangle|^2 [ 1 - F_N(\omega_{\nu\mu})] + \nonumber\\ \Gamma_N |\langle \nu|A_1 +
A_2|\mu\rangle|^2 F_N(\omega_{\mu\nu}),
\end{eqnarray}
and by a similar rate $\gamma_{\mu\nu}^{PQ}$. Here
\begin{equation}
F_{\sigma}(E) = \left[\exp\left(\frac{E-\mu_{\sigma}}{T}\right) + 1\right]^{-1}
\end{equation}
is the Fermi distribution of the protons in $\sigma-$reservoir ($\sigma = N, P$).

The average population of the A-site is governed by the equation
\begin{eqnarray} \label{nA}
\langle \dot{n}_A \rangle = \gamma_S \,[ f_S(\varepsilon_A) - \langle n_A \rangle ] +  \nonumber\\ |\Delta_{AQ}|^2
\sqrt{\frac{\pi}{\lambda_{AQ}T }}\; \sum_{\mu\nu} |\langle \mu|a_1+a_2|\nu\rangle |^2 \times \nonumber\\ \left\{  \langle 1 - n_A \rangle
\exp\left[ -\frac{(\omega_{\mu\nu} + \varepsilon_A + \lambda_{AQ})^2}{4\lambda_{AQ}T}\right] \langle \rho_{\nu}\rangle \right. \nonumber\\
\left. - \langle n_A \rangle \exp\left[ -\frac{(\omega_{\mu\nu} + \varepsilon_A - \lambda_{AQ})^2}{4\lambda_{AQ}T}\right] \langle
\rho_{\mu}\rangle \right\},
\end{eqnarray}
where a coefficient $\gamma_S$ (or $\gamma_D$) describes the electron transitions from the A (or B) site to the source S (drain D) electron
reservoir characterized by a Fermi distribution with electrochemical potentials $\mu_S$ or $\mu_D$ ($\alpha = S,D$),
\begin{equation}
f_{\alpha}(\varepsilon) = \left[\exp\left(\frac{\varepsilon - \mu_{\alpha}}{T}\right) + 1\right]^{-1}.
\end{equation}
A similar equation takes place for the population $\langle n_B \rangle$.

We solve the rate equations (\ref{RMN},\ref{rhoMuNu},\ref{nA}) for both the distributions $\langle R_{M}\rangle,\, \langle \rho_{\mu}\rangle$
and for the populations $\langle n_A\rangle$ and $ \langle n_B \rangle$, together with an overdamped Langevin equation for the mechanical motion
of the shuttle,
\begin{eqnarray}
\zeta\, \dot{x} \,=\, -\, \frac{d U_{\rm w}}{dx} - \langle (n_1 + n_2 - N_1 - N_2)^2\rangle \, \frac{d U_{\rm ch}}{dx} + \xi,
\end{eqnarray}
where $\zeta$ is the drag coefficient, $\xi$ is a Gaussian fluctuation source with zero mean value, $\langle \xi \rangle = 0$, and with a
correlator $\langle \xi(t) \xi(t')\rangle = 2 \zeta T \delta (t-t').$ The potential $U_{\rm w}(x)$ confines the shuttle between the membrane
walls, and the potential $U_{\rm ch}(x)$ prevents the charged molecule Q from crossing the lipid core of the membrane (for details see
Ref.~\cite{SmirnovPRE09} ). We note that the tunneling amplitudes $\Delta_{AQ}, \Delta_{HQ}$ and the proton rate $\Gamma_N$ depend on the
distance between the shuttle (with a coordinate $x$) and the N-side catalytic center located at $x = - x_0,$ whereas the tunneling amplitudes
$\Delta_{BQ}, \Delta_{LQ}$ and the proton rate $\Gamma_P$ depend on the distance between the shuttle and the P-side catalytic center located at
$x = x_0$.

\end{appendix}

\newpage


\newpage

\begin{figure}
\includegraphics[width=12.0cm, height=15.0cm ]{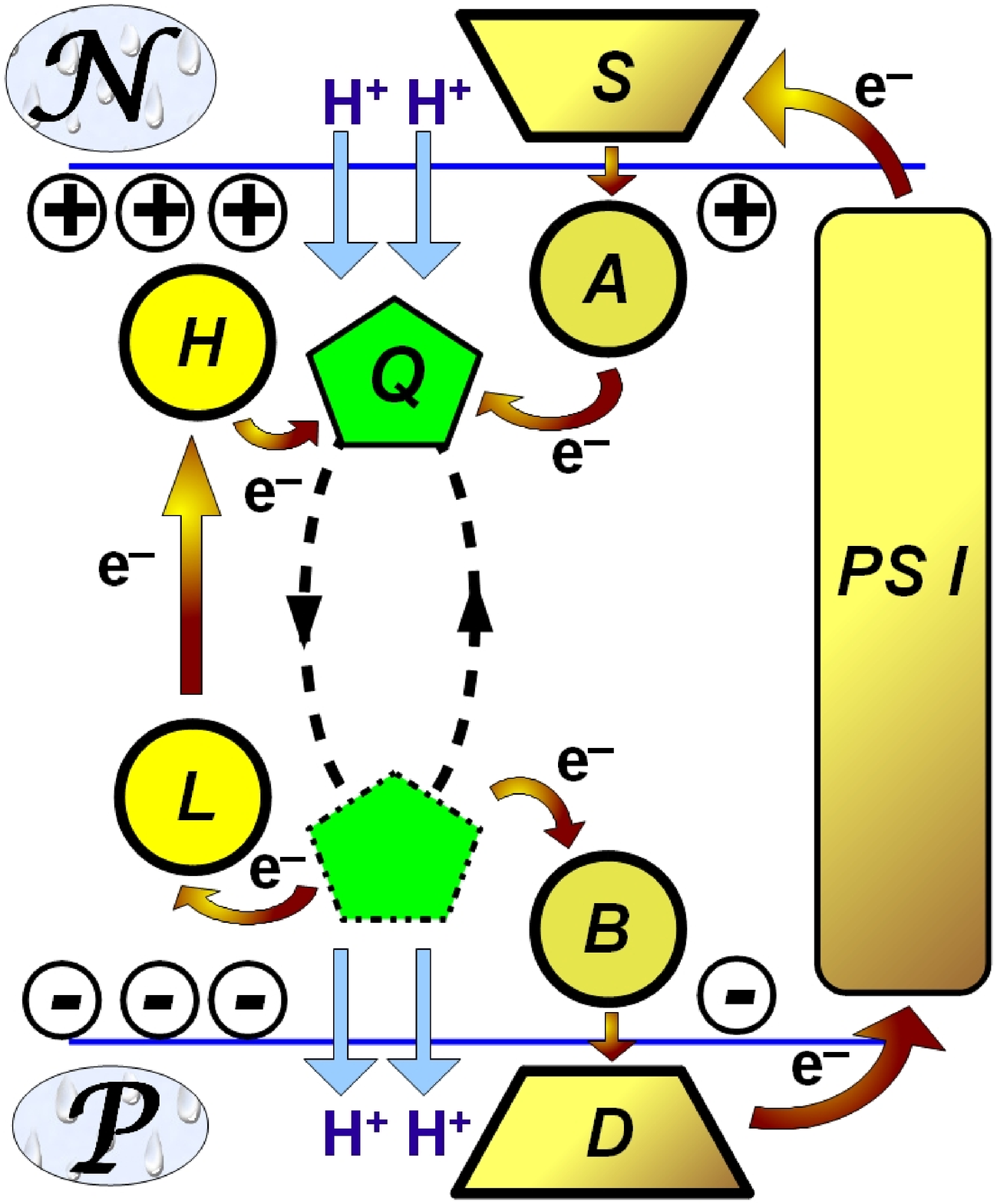}
\vspace*{0cm} \caption{ Simplified schematic of the Q-cycle mechanism in the regime of cyclic electron flow. At the N-side of the membrane the
Photosystem I (PS I) provides the source reservoir S with high-energy electrons. Via the bridge site A, the source S delivers electrons to the
shuttle Q, which also accepts electrons from the site H and protons from the N-side proton reservoir. At the P-side of the membrane the shuttle
Q gives away electrons to the site L and to the drain reservoir D (via the bridge site B). In this process, two protons move to the P-side
proton reservoir. From the site L electrons return to the site H to be loaded later on the shuttle Q. The drain reservoir D transfers low-energy
electrons back to the Photosystem I. The surface potential $V_S$, which is positive at the N-side and negative at the P-side of the membrane, is
shown here with circled plus and minus signs.}
\end{figure}

\begin{figure}
\includegraphics[width=20.0cm, height=12.0cm ]{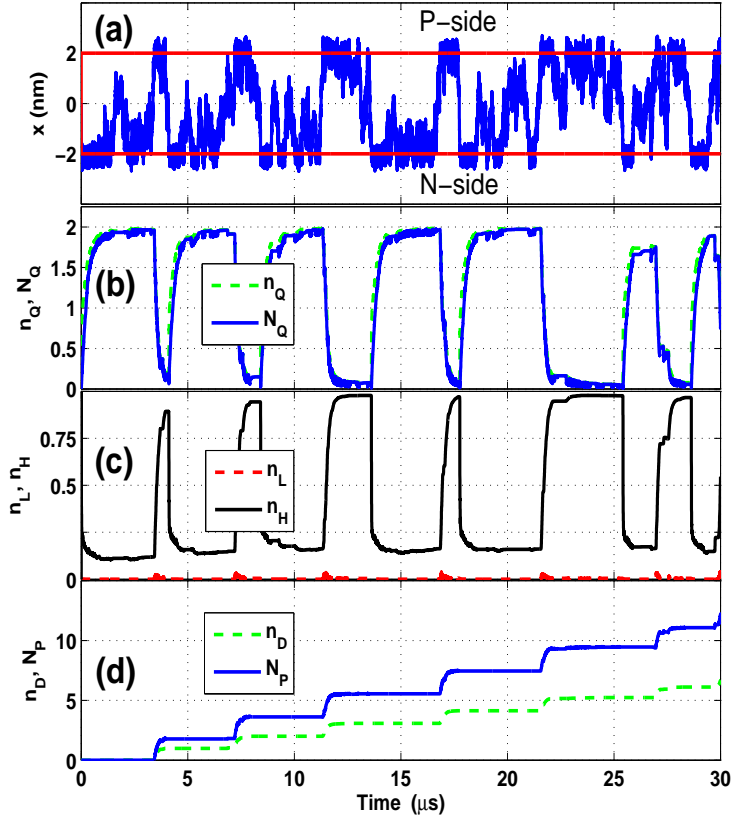}
\vspace*{0cm} \caption{  (a) Time evolution of the position $x(t)$ of the shuttle diffusing between the membrane walls located at $x = \pm
2$~nm; (b)-(d) a sequence of electron and proton transfer reactions at the following set of parameters: $\mu_S - \mu_D = 850, \, \mu_P - \mu_N =
150, \, U_{ep} = 610$~meV, and $\Delta V = 260$~meV. The total proton population $N_Q$ of the shuttle [blue continuous line in (b)] almost
coincides with the electron population $n_Q$ marked by the dashed green line. The bifurcated electron transfer reaction takes place at the
P-side of the membrane (at $x = 2$~nm) where one electron moves from Q to the site B and the drain D [see, e.g., a step down for $n_Q$ in (b)
and a step up for $n_D$ in (d) at the moment $t \sim 3~\mu$s]. At almost the same time another electron moves to the site L [see a barely
visible dashed red line in (c)] and rapidly proceeds to the site H [blue continuous spike in (c)]. Two protons are unloaded from the shuttle to
the P-side of the membrane as follows from the step for $N_P$ [blue continuous curve in (d)]. It can be seen from (d) that the number of protons
$N_P$ translocated to the P-side (blue continuous line) is nearly twice as large as the number of electrons $n_D$ moved to the drain (dashed
green line).}
\end{figure}

\begin{figure}
\includegraphics[width=20.0cm, height=12.0cm ]{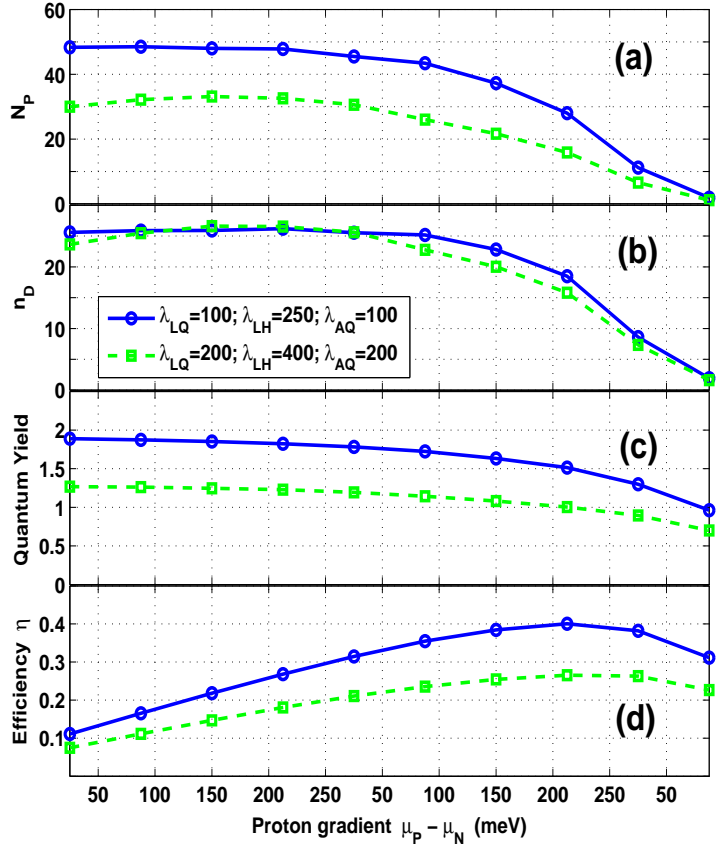}
\vspace*{0cm} \caption{ (a) Number of protons $N_P$ transferred to the P-side of the membrane; (b) number of electrons $n_D$ moved to the drain;
(c) their ratio (quantum yield $QY$); (d) the thermodynamic efficiency $\eta$ as functions of the transmembrane proton gradient $\mu_P - \mu_N$
(measured in meV) for different couplings to the environment (in meV): weaker couplings (i) $\lambda_{LQ} = 100;\,\lambda_{LH} = 250;\,
\lambda_{AQ}=100$ (blue continuous curves); and stronger couplings (ii)  $\lambda_{LQ} = 200;\,\lambda_{LH} = 400;\, \lambda_{AQ}=200$ (green
dashed curves). Other parameters are the same as in Fig.~2. The data are averaged over 10 stochastic realizations. Each realization lasts for
100 $\mu$s. These graphs demonstrate the ability of the system to translocate protons against the gradient up to 200~meV with an efficiency
$\eta $ up to 40\%.}
\end{figure}

\begin{figure}
\includegraphics[width=20.0cm, height=12.0cm ]{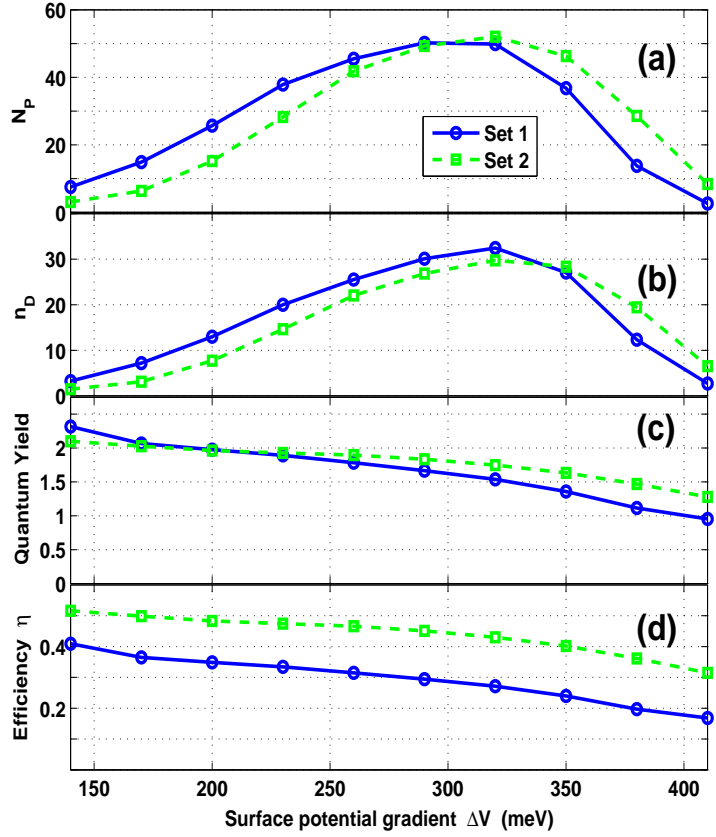}
\vspace*{0cm} \caption{ Dependence of the output indicators of the system, the number of protons translocated to the P-side, $N_P$ (a), the
number of electrons moved to the drain, $n_D$ (b), the quantum yield $QY (c)$ and the efficiency $\eta$ (d), on the surface potential gradient
$\Delta V = V_N + V_P$ across the membrane. Two kinds of curves correspond to two sets of energy values: (1) $\mu_S - \mu_D = 850;\, \mu_P -
\mu_N = 150;\, U_{ep} = 610$~meV (blue continuous curves), and (2) $\mu_S - \mu_D = 1220;\, \mu_P - \mu_N = 300;\, U_{ep} = 800$~meV (green
dashed curves) and for $\lambda_{LQ} = 100,\,\lambda_{LH} = 250,\, \lambda_{AQ}=100$~meV. It can be seen from this figure that the system is
operational in the range of the surface potentials from $\Delta V = 230$~meV up to $\Delta V = 350$~meV. In agreement with Eq.~(\ref{EqEta}) the
complex works more efficiently (with $\eta \sim 50\%$ and $QY \sim 2$) for the second set of parameters where both electron and proton gradients
are higher. }
\end{figure}

\end{document}